# All-Optical Excitation and Detection of Picosecond Dynamics of Ordered Arrays of Nanomagnets with Varying Areal Density

Bivas Rana,[1] Semanti Pal,[1] Saswati Barman,[1] Yasuhiro Fukuma,[2] YoshiChika Otani[2,3] and Anjan Barman[*,1]

[1]*Department of Material Sciences, S. N. Bose National Centre for Basic Sciences, Block JD, Sector III, Salt Lake, Kolkata 700 098*

[2]*RIKEN ASI, 2-1 Hirosawa, Wako, Saitama 351-0198, Japan*

[3]*Institute for Solid State Physics, University of Tokyo, 5-1-5 Kashiwanoha, Kashiwa, Chiba 277-8581, Japan*

We have demonstrated optical excitation and detection of collective precessional dynamics in arrays of coupled $Ni_{80}Fe_{20}$ (permalloy) nanoelements with systematically varying areal density by an all-optical time-resolved Kerr microscope. We have applied this technique to precisely determine three different collective regimes in these arrays. At very high areal density, a single uniform collective mode is observed where the edge modes of the constituent elements are suppressed. At intermediate areal densities, three nonuniform collective modes appear and at very low areal density, we observe noncollective dynamics and only the centre and edge modes of the constituent elements appear.

*E-mail address: abarman@bose.res.in

Ordered arrays of nanomagnets have inspired technological progress within information storage,[1] memory,[2] spin logic[3], spin torque nanooscillators (STNOs),[4] and magnonic crystals.[5,6] Magnetostatically coupled nanomagnets in an ordered array may show long wavelength collective dynamics,[7-9] where the dynamics of the constituent nanomagnets maintain a constant amplitude and phase relationships similar to the acoustic and optical modes of phonons. Such long wavelength collective modes pose obstacles to the applications of nanomagnet arrays in storage, memory devices, and magnetic field sensors, where the individual characters of the nanomagnets (bits) must be retained. On the other hand, to achieve a more useful power level, a microwave emitter should consist of arrays of phase coherent nano-oscillators. The long wavelength collective dynamics in the form of Bloch waves, defined in the first Brillouin zone (BZ) of an artificial lattice, can also be manipulated by tailoring the lattice to form two-dimensional magnonic crystals.[10,11]

Magnetization dynamics in planar arrays of nanomagnets and single nanomagnets have been experimentally studied by time-domain,[7,9,10,12,13] frequency-domain,[14] and wave-vector-domain[8,11,15] techniques and by analytical methods[16,17] and micromagnetic simulations.[18] To this end the frequency, damping, and spatial patterns of collective modes and dispersion relations of frequency with the wave-vector of magnon propagation have been studied. However, very few attempts have been made to understand the systematic variation of collective dynamics in arrays with varying areal density. This is even more important when the constituent elements possess nonuniform magnetic configurations. The collective dynamics is determined by the interplay between the static and dynamic stray fields and very little is known about the effects of the complex stray field and magnetic ground states of the individual elements on the collective dynamical modes in an array. Here, we demonstrate the excitation and detection of collective precessional dynamics in arrays of square permalloy elements with 200 nm width and with varying interelement separation by an all-optical time-resolved magneto-optical Kerr effect microscope. Our technique enables us to avoid complicated sample fabrication process on waveguide structures and to obtain a much better temporal resolution (~100 fs) than the nonoptical excitation techniques. We observe a systematic variation from a single frequency uniform collective precession to a number of nonuniform collective modes and finally to isolated dynamics of the single elements with the decrease in areal density. A concurrent variation of the precessional modes is observed with the variation of the bias magnetic field confirming the importance of the magnetic ground states of the whole array. We interpret the experimental observations with the aid of micromagnetic simulations.

The samples used in this experiment are $10 \times 10$ $\mu m^2$ square arrays of permalloy elements with 200 nm width, 20 nm thickness, and interelement (edge to edge) separation ($S$) varying from 50 to 400 nm. The samples are prepared by electron beam evaporation of permalloy on prepatterned substrates prepared by electron-beam lithography followed by a lift-off technique. Some residual resists remained on top and at the sides of the patterned elements, which were further removed by oxygen plasma. Figure 1(a) presents the scanning electron micrographs of the arrays, which show that the arrays were well fabricated with < 5% deviation from the nominal dimensions. A permalloy square element with 10 μm width and 20 nm thickness was also prepared to obtain the magnetic parameters of the unpatterned sample. The ultrafast magnetization dynamics was measured using a custom-built time-resolved magneto-optical Kerr effect microscope based upon a two-color all-optical pump-probe setup in a collinear manner.[19] The second harmonic ($\lambda$ = 400 nm) of a Ti-sapphire laser (Tsunami, SpectraPhysics, pulse-width < 80 fs) was used to pump the samples, while the time-delayed fundamental ($\lambda$ = 800 nm) laser beam was used to probe the dynamics by measuring the polar Kerr rotation by means of a balanced photodiode detector, which completely isolates the Kerr rotation and the total reflectivity signals. The pump and probe beams were focused and spatially overlapped at the centre of each array by a microscope objective with numerical aperture N. A. = 0.65 in a collinear geometry. A large magnetic field is first applied at a small angle (~15°) to the sample plane to saturate its magnetization. The magnetic field strength is then reduced to the bias field value ($H$ = component of bias field along the x-direction), which ensures that the magnetization remains saturated along the bias field direction. The pump beam was chopped at 2 kHz frequency and a phase sensitive detection of the Kerr rotation was used. The precessional dynamics appears as an oscillatory signal above the slowly decaying part of the time-resolved Kerr rotation after a fast demagnetization within 500 fs, and a fast remagnetization within 10 ps. A bi-exponential background is subtracted from the time-resolved signal before performing the fast Fourier transform (FFT) to find out the corresponding power spectra.

Figure 1(b) shows the precessional frequency as a function of $H$ measured from the centre of the 10 × 10 μm² permalloy element. The experimental data is fitted with the Kittel's formula for the uniform precession of magnetization and the magnetic parameters were obtained as $\gamma$ = 18.5 MHz/Oe, anisotropy field $H_K$ = 0, and saturation magnetization $M_S$ = 860 emu/cc. Figure 2(a) shows the experimental time-resolved Kerr rotations from the arrays of permalloy elements with varying $S$ at $H$ = 1.25 kOe. The corresponding FFT spectra are shown in Fig. 2(b). A clear variation in the precession frequencies is observed with the increase in $S$. For $S$ = 50 and 75 nm, a single resonant mode is primarily observed. For $S$ = 75 nm the precession decays faster than that for $S$ = 50 nm indicating the appearance of an inhomogeneous line broadening. For $S$ = 100 and 150 nm, clear broadening and partial splitting of the resonance mode is observed in addition to a lower frequency mode. For $S$ > 150 nm, three distinct modes are consistently observed but at $S$ = 400 nm, only two clear modes are observed. In Fig. 2(c), we show the results obtained from micromagnetic simulations of arrays of 7 × 7 square permalloy elements using OOMMF software.[20] The material parameters used in the simulation are $\gamma$ = 18.5 MHz/Oe, $H_K$ = 0, and $M_S$ = 860 emu/cc as obtained from the Kittel fit described above, while the exchange stiffness constant A = 1.3 × 10⁻⁶ erg/cm is obtained from the literature.[21] In the simulation, we introduced a ±5% deviation in the width and interelement separation to mimic the real samples. The simulations qualitatively reproduce the key features of experimental observation including the appearance of a single resonance mode at $S = $ 50 and 75 nm, broadening and splitting of the resonant mode at $S = $ 100 and 150 nm, and the appearance of a number of modes for $S > $ 150 nm. In Fig. 3, we present the bias magnetic field dependence of the precessional dynamics for the array with $S = $ 50 nm. The frequencies of the peaks reduce and the single resonance peak gets broadened and consequently breaks into two or more modes with the decrease in $H$. The decay rate of precession also increases with the decrease in bias field but we did not attempt to quantify the damping coefficient of precession due to the incoherent precession for the lower bias fields. In Fig. 3(b) we plot the precession frequencies as a function of bias magnetic field. The highest frequency peak fits well with the Kittel mode calculated with the parameters obtained for the $10 \times 10$ μm$^2$ permalloy element, showing that this mode corresponds to the collective precession of all elements in the $10 \times 10$ μm$^2$ array. At lower bias fields, a lower frequency mode appears whose frequency also systematically decreases with the bias field. Micromagnetic simulation of the bias field dependence of the precessional dynamics of the array with $S = $ 50 nm reproduces the experimental results qualitatively. However, at lower bias fields the higher frequency mode splits into two closely spaced modes in the simulation, which was not resolved experimentally. Furthermore, the lower frequency mode, observed experimentally, does not quantitatively agree with the experimental data possibly due to the physical differences between the experimental and simulated samples.

Figure 4(a) shows the spatial maps of the x-component of simulated magnetostatic field distribution from the central part of the arrays (3 × 3 elements) with $S$ = 50, 100, and 300 nm. The stray magnetic field is very high (~ 2 kOe at the centre of the gaps between the elements) for $S$ = 50 nm and decreases steeply with an increase in $S$ and becomes < 200 Oe for $S$ = 300 nm. This causes a concurrent reduction in the coherence between the precession of the constituent elements in the arrays. The spatial characters of the observed modes are simulated by resonantly exciting the arrays with ac fields.[22] The spatial maps of magnetization are calculated after the spurious nonresonant modes decay naturally, leaving only the driven resonant mode. The magnetization maps of the resonant modes from the central part of the arrays (3 × 3 elements) with $S$ = 50, 100, and 300 nm are shown in Fig. 4(b), along with the simulated centre and edge modes of the isolated 200 × 200 nm$^2$ element.[22,23] For $S$ = 50 nm, we observe a collective precession of magnetization of all elements in the array, which is visibly different from the centre mode of the individual elements. A small amount of nonuniformity near the edges is observed, possibly due to the complex profile of the magnetostatic fields inside the arrays. For $S$ = 100 nm, the highest frequency mode (peak 1) corresponds to the collective precession of the elements similar to the sole mode observed for the array with $S$ = 50 nm. The lower frequency mode (peak 2) has a spatial profile similar to the backward volume magnetostatic (BWVMS) modes. Peak 3 corresponds to a localized mode near the edges of the elements. For the array with $S$ = 300 nm (width/separation < 1) the interaction between the elements is very weak and peaks 1 and 2 correspond to the centre and edge modes of the 200 nm wide element, respectively. However, peak 3 is possibly a long wavelength collective mode, which was not observed in the isolated element. At $S$ = 400 nm, only the centre and edge modes of the isolated elements are observed in all elements of the array.

In summary, we have demonstrated that collective dynamics in arrays of nanomagnets can be excited and detected in an all-optical manner. By using this technique, we studied the systematic variation of collective precessional dynamics as a function of the areal density of the arrays. The stray magnetic field is very high for higher areal density, which systematically decreases with the decrease in areal density. Consequently, the dynamics show a single frequency collective precession of all elements in the array for very high areal density, where the edge modes of the constituent elements are suppressed. With the decrease in areal density we observe the onset of nonuniform collective modes of the array. At very low areal density, the elements become uncoupled and centre and edge modes of the isolated elements dominate. The precise determination of the length scales corresponding to the transition between various collective regimes is important for the applications of these arrays in various nanomagnet-based devices.

The authors gratefully acknowledge the financial support from the Department of Science and Technology, Government of India under grant numbers SR/NM/NS-09/2007, INT/EC/CMS (24/233552), and INT/JP/JST/P-23/09 and Japan Science and Technology Agency Strategic International Cooperative Program under grant number 09158876.

**Figure Captions**

Fig. 1. (a) Scanning electron micrographs of arrays of permalloy square elements with width = 200 nm, thickness = 20 nm, and varying interelement separation $S$. The experimental geometry is shown on top of the image. (b) Experimental precession frequency (symbols) is plotted as a function of $H$ for a square permalloy element with width = 10 µm and thickness = 20 nm. The solid line corresponds to the fit with Kittel's equation.

Fig. 2. (a) Experimental time-resolved Kerr rotations and (b) the corresponding FFT spectra are shown for arrays of permalloy square elements with width = 200 nm, thickness = 20 nm, and varying interelement separation $S$ at $H = 1.25$ kOe. (c) FFT spectra of the simulated time-resolved magnetization..

Fig. 3. (a) Experimental time-resolved Kerr rotations and the corresponding FFT spectra are shown for the array with $S = 50$ nm at different $H$. (b) Precession frequencies (filled symbols: experimental data, hollow symbols: micromagnetic simulation, solid line: Kittel fit) are plotted as a function of $H$.

Fig. 4. (a) Simulated magnetostatic field distributions (x-component) are shown for arrays of square permalloy elements with $S = 50$, 100, and 300 nm at $H = 1.25$ kOe. The dotted lines show the physical outlines of the elements in the arrays. The simulated spatial profile of different precessional modes of (b) arrays of permalloy elements and (c) an isolated permalloy element are shown.

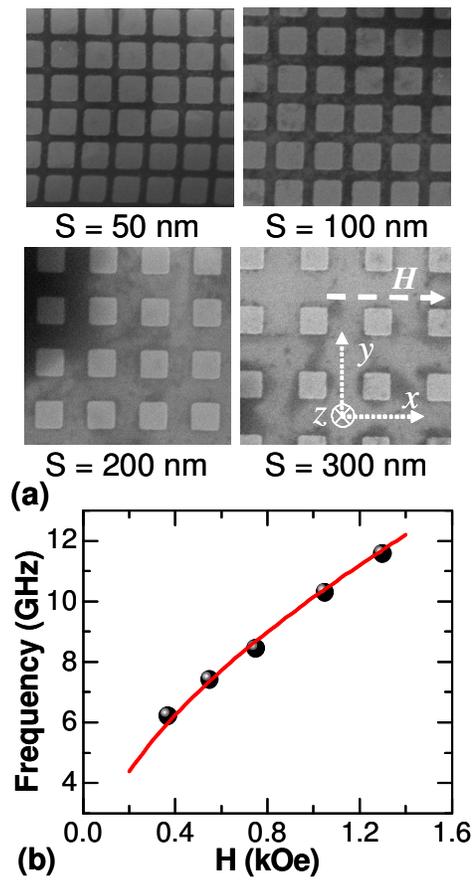

Fig. 1

B. Rana et al.

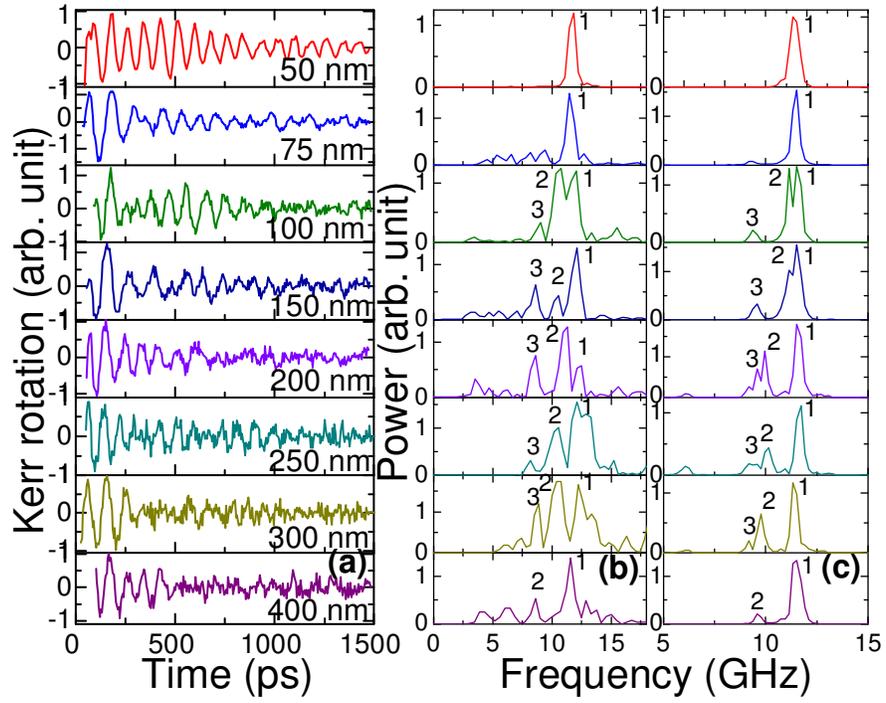

Fig. 2

B. Rana et al.

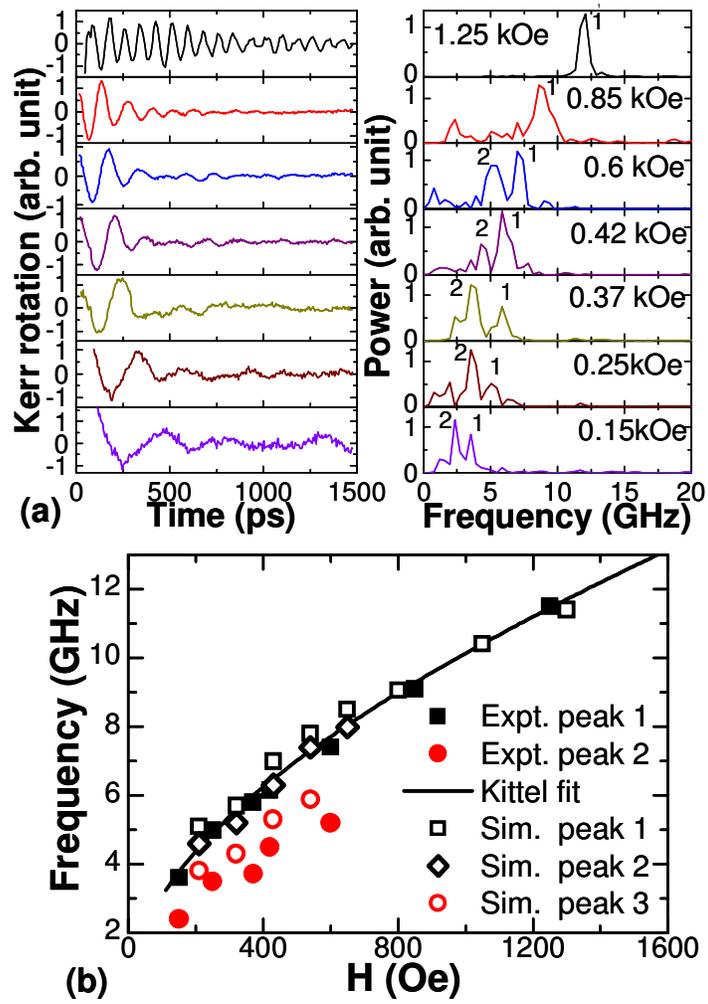

Fig. 3

B. Rana et al.

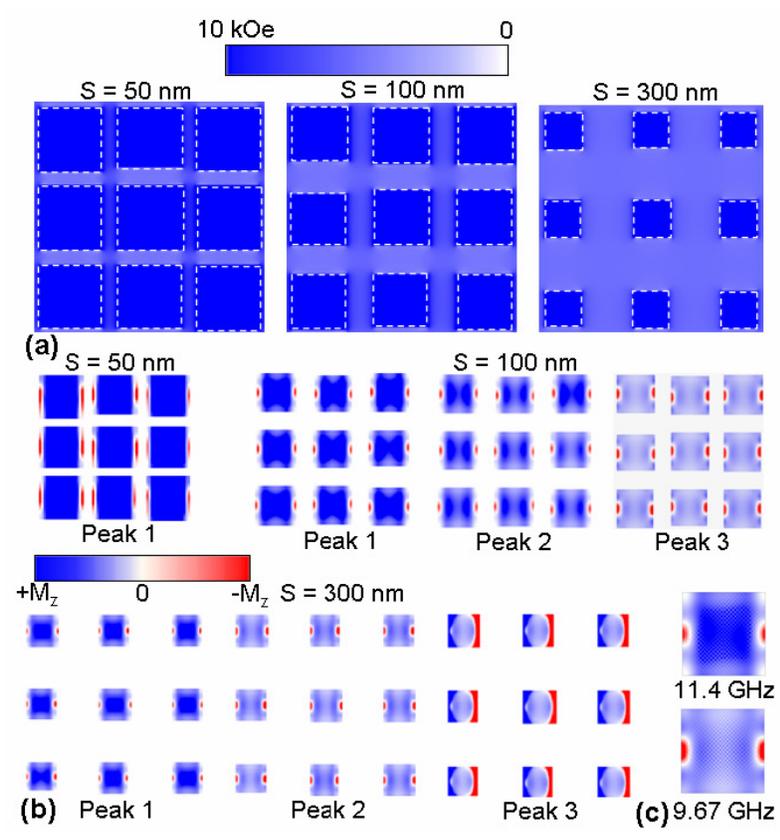

Fig. 4

B. Rana et al.